\documentclass[aps,prb,twocolumn,floatfix,amsmath,amssymb,showpacs,
               superscriptaddress]{revtex4}
\usepackage{graphicx}
\usepackage{color}
\usepackage{dcolumn}
\usepackage{bm}
\usepackage{epsfig,psfrag,amsmath,amssymb,float}
\input{epsf}

\begin{document}
\title{Magnetic and pairing properties of a two-orbital model for the \\
pnictide superconductors: A quantum Monte Carlo study}
\author{Guang-Kun Liu}
\email{gkliu@mail.bnu.edu.cn}
\affiliation{Department of Physics,
Beijing Normal University, Beijing 100875, China}

\author{Zhong-Bing Huang}
\email{huangzb@hubu.edu.cn}
\affiliation{Department of Physics, Hubei University,
Wuhan 430062, China\\}
\affiliation{Beijing Computational Science Research Center,
Beijing 100084, China}

\author{Yong-Jun Wang}
\email{yjwang@bnu.edu.cn}
\affiliation{Department of Physics, Beijing Normal University,
Beijing 100875, China}

\date{\today}

\begin{abstract}
Using the constrained-path Monte Carlo method, a two-orbital model
for the pnictide superconductors is studied at half filling and in
both the electron- and hole-doped cases. At half filling, a stable
$(\pi,0)$/$(0,\pi)$ magnetic order is explicitly observed, and the
system tends to be in an orthomagnetic order rather than the striped
antiferromagnetic order when increasing the Coulomb repulsion $U$.
In the electron-doped case, the $(\pi,0)$/$(0,\pi)$ magnetic order
is enhanced upon doping and suppressed eventually, and a $s_{\pm}$
pairing state dominates all the possible nearest-neighbor-bond
pairings. Whereas in the hole-doped case, the magnetic order is
straightforwardly suppressed and two nearly degenerate $A_{1g}$ and
$B_{1g}$ intraband pairings become the dominant ones.
\end{abstract}


\maketitle
%%the introduction part
\section{Introduction}
%%reaserch background:list the main problem
The discovery of pnictide superconductors (SCs) has triggered lots
of attentions of the condensed matter community. Unlike cuprates,
where only the Cu $3d_{x^2-y^2}$ orbital plays the most significant
role, the local-density approximation (LDA)
calculations\cite{Vildosola2008,Mazin2008} indicate that pnictide
SCs have several active $3d$ orbitals near the Fermi surface (FS).
Consequently, it is widely believed that such SCs should be
understood in terms of multi-orbital models instead of the
single-orbital ones.\cite{Raghu2008,Daghofer2010b,Kuroki2008}
Regarding the minimal model capable of capturing the essential
physics of pnictide SCs, some authors proposed more realistic three-
and five-orbital models,\cite{Daghofer2010b,Kuroki2008} while others
argued that the main physics of the pnictide SCs are contained in
two-orbital models.\cite{Moreo2009,Raghu2008,Hu2012} Because of
their relative simplicity, as well as the fact that\cite{Moreo2009}
the correct FS shape can be reproduced in both the doped and undoped
cases, it is crucial to find out the properties of these two-orbital
models.

Most previous theoretical works on the two-orbital models for
pnictides were based on mean-field-like approximations, such as the
random phase approximation (RPA), fluctuation exchange (FLEX) and
functional renormalization group (fRG) calculations and so on. The
main results from these studies provided good understandings of
pnictides while still giving very different pairing pictures. For
instance, Graser \emph{et al.} proposed two nearly degenerate
competing pairing states with $A_{1g}$ and $B_{1g}$ symmetries
because of the near nesting of FS sheets.\cite{Graser2009} Although
other RPA studies also suggested a competing pairing picture, different
pairing channels were proposed, for example, the competition between
singlet $d$-wave and triplet $p$-wave states\cite{Qi2008} or $s_{\pm}$-wave
and $d$-wave states.\cite{Bang2008} Aside from RPA studies, a FLEX calculation
demonstrated an $s_{\pm}$-wave or $d_{xy}$-wave pairing depending on
whether the intraband antiferromagnetic (AFM) spin fluctuation is
stronger than the interband one or not\cite{Yao2009}. Meanwhile, fRG
approach revealed $s_{\pm}$-wave and sub-dominant $d$-wave
pairings.\cite{Wang2008} Moreover, Dai \emph{et al.} suggested a
spin triplet pairing state by using of the BCS mean field
method.\cite{Dai2008}

From the above discussions, it seems of great difficulty to justify
the pairing symmetry of the two-orbital models. Since it is
unrealistic to take the full quantum fluctuations into account in
the usual theoretical methods, the important role of electronic
correlations on the magnetism and superconductivity has not been
thoroughly recognized. Despite the fact that there exist some
unbiased numerical investigations of the two-orbital
model,\cite{Daghofer2008,Nicholson2011,Moreo2009,Berg2010} these
results are not sufficient enough to understand the electronic
correlations in the two-dimensional systems because they are
obtained either on a 8-site
lattice\cite{Daghofer2008,Nicholson2011,Moreo2009} or on a diagonal
ladder.\cite{Berg2010} For a comprehensive and systematic
understanding of the two-orbital model,\cite{Daghofer2008,Raghu2008}
a quantum Monte Carlo (QMC) method is employed in this paper.

However, it is known that there are several difficulties in the QMC
simulations of the two-orbital models: One is the severe limitation
of the cluster size. For example, the computational demanding of a
two-orbital model with a given cluster is much higher than that
of an one-orbital model with a double cluster size.
Another\cite{Held1998,Sakai2004}, probably the most tough one, is the
insurmountable Fermi sign problem. Compared to the one-orbital model,
a usual discrete Hubbard-Stratonovich (HS) transformation\cite{Hirsch1983}
for the Hund's coupling and pair-hopping terms, which is
specific to the multi-orbital models, does lead to a more serious
sign problem. It is noticed that Sakai~\emph{et al.}\cite{Sakai2004}
proposed a new type of transformation for the Hund's coupling and
pair-hopping interaction, which can effectively alleviate the sign problem.
Based on this progress, we developed a feasible constrained-path Monte
Carlo (CPMC) method~\cite{Zhang1997} for the two-orbital models,
which works well in the weak and intermediate correlation regimes.

Using the CPMC method, we compute the magnetic structure
factors and the pairing correlations of the two-orbital model as
functions of the doping density $\rho$ and interaction strength $U$.
We find a ($\pi,0$)/($0,\pi$) magnetic order that is enhanced by the
Coulomb repulsion $U$ and the Hund's coupling $J$ in the undoped
case. Because of the particle-hole asymmetry of the two-orbital model,
such a magnetic order shows different behaviors in the electron- and
hole-doped cases. We also find that the doping has much stronger effect
than that of the Coulomb repulsion on the pairing correlations.
In the electron-doped case, a nodeless $s_{\pm}$ pairing is dominant,
whereas in the hole-doped case, two nearly degenerate $A_{1g}$ and $B_{1g}$
intraband singlet pairings compete with each other and become the
dominant ones.

Our paper is organized as follows: In Sec.~\ref{2nd}, we briefly
introduce the two-orbital model under investigation and discuss the
proper choice of model parameters. Some modifications to the
original CPMC algorithm and the definitions of the calculated
physical quantities are presented as well. In Sec.~\ref{3rd}, we
exhibit in details the simulation results for the magnetic and
pairing properties of the model with various parameters. Our main
conclusions are summarized in Sec.~\ref{4th}.

%%the second part
\section{MODEL AND NUMERICAL APPROACH}\label{2nd}
%% simple description of the model
   On the basis of LDA calculations, Mazin \emph{et al.}
\cite{Mazin2008} advocated that the band structure of pnictides
involves only three Fe $3d$ orbitals, $d_{xz}$, $d_{yz}$, and
$d_{xy}$ (or $d_{x^2-y^2}$), near the Fermi level. Accordingly,
Raghu \emph{et al.}\cite{Raghu2008} introduced a minimal
multi-orbital model for the pnictide SCs within the further
approximation that a next-near-neighbor hybridization between
$d_{xz}$, $d_{yz}$ orbitals can be equated to the role of the
$d_{xy}$ or $d_{x^2-y^2}$ orbital. As described in
Ref.~\onlinecite{Daghofer2008}, the kinetic part of the
two-orbital model Hamiltonian is given by
\begin{eqnarray}\label{eq1}
&& H_0=
-t_1\sum_{\textbf{i},\sigma}(d_{\textbf{i},x,\sigma}^{\dagger}d_{\textbf{i}+\hat{y},x,\sigma}+
                         d_{\textbf{i},y,\sigma}^{\dagger}d_{\textbf{i}+\hat{x},y,\sigma}+\mathrm{h.c.}) \notag \\
    &&\;\;-t_2\sum_{\textbf{i},\sigma}(d_{\textbf{i},x,\sigma}^{\dagger}d_{\textbf{i}+\hat{x},x,\sigma}+
                         d_{\textbf{i},y,\sigma}^{\dagger}d_{\textbf{i}+\hat{y},y,\sigma}+\mathrm{h.c.}) \notag \\
    &&\;\;-t_3\sum_{\textbf{i},\hat{\mu},\hat{\nu},\sigma}(d_{\textbf{i},x,\sigma}^{\dagger}
                                                       d_{\textbf{i}+\hat{\mu}+\hat{\nu},x,\sigma}+
                                                       d_{\textbf{i},y,\sigma}^{\dagger}
                                                       d_{\textbf{i}+\hat{\mu}+\hat{\nu},y,\sigma}+
                                                         \mathrm{h.c.}) \notag \\
    &&\;\;+t_4\sum_{\textbf{i},\sigma}(d_{\textbf{i},x,\sigma}^{\dagger}d_{\textbf{i}+\hat{x}+\hat{y},y,\sigma}+
                 d_{\textbf{i},y,\sigma}^{\dagger}d_{\textbf{i}+\hat{x}+\hat{y},x,\sigma}+\mathrm{h.c.}) \notag \\
    &&\;\;-t_4\sum_{\textbf{i},\sigma}(d_{\textbf{i},x,\sigma}^{\dagger}d_{\textbf{i}+\hat{x}-\hat{y},y,\sigma}+
                         d_{\textbf{i},y,\sigma}^{\dagger}d_{\textbf{i}+\hat{x}-\hat{y},x,\sigma}+\mathrm{h.c.})
                         ,
\end{eqnarray}
where $x$ and $y$ represent the $d_{xz}$ and $d_{yz}$ orbitals,
respectively. The operator $d_{\textbf{i}\alpha\sigma}^\dagger$
creates an electron on orbital $\alpha$ in Fe site $\textbf{i}$ with
spin $\sigma$, and the index $\hat{\mu}(\hat{\nu})=\hat{x}$ or
$\hat{y}$ denotes a unit vector linking the nearest-neighbor sites.
To estimate the hopping amplitudes that can recover the right
topology of Fermi surface and band features given by
DFT,\cite{Singh2008,Xu2008} the band-structure
calculation\cite{Raghu2008} and the Slater-Koster tight-binding
scheme\cite{Moreo2009} recommended different hopping amplitudes,
however, the Lanczos study on 8-site cluster suggested these two
schemes give similar physics\cite{Moreo2009}. Following the
band-structure calculation, the hopping parameters will always be
taken as $t_1=-1.0$, $t_2=1.3$ and $t_3=t_4=-0.85$ in our
calculation.

 The interaction terms,\cite{Daghofer2008,Bascones2010,Luo2010} containing a Hubbard
repulsion in the same orbital, a repulsion $U^{\prime}$ for
different orbitals, a ferromagnetic Hund's coupling $J$, and
pair-hopping terms, can be expressed as
\begin{equation}\begin{split}\label{eq2}
&H_{int}=\sum_{\textbf{i}}(H_1^\textbf{i}+H_2^\textbf{i}+H_3^\textbf{i}+H_4^\textbf{i}),\\
&H_1^\textbf{i}=J\sum_{\alpha\neq\alpha^\prime}
       (d_{\textbf{i}\alpha\uparrow}^\dagger d_{\textbf{i}\alpha^\prime\downarrow}^\dagger
        d_{\textbf{i}\alpha\downarrow} d_{\textbf{i}\alpha^\prime\uparrow}\\
       &\hspace{2.5em} +d_{\textbf{i}\alpha\uparrow}^\dagger d_{\textbf{i}\alpha\downarrow}^\dagger
        d_{\textbf{i}\alpha^\prime\downarrow} d_{\textbf{i}\alpha^\prime\uparrow}),\\
&H_2^\textbf{i}=(U^\prime-J)\sum_{\sigma}n_{\textbf{i},x,\sigma}n_{\textbf{i},y,\sigma},\\
&H_3^\textbf{i}=U\sum_{\alpha}n_{\textbf{i}\alpha\uparrow}n_{\textbf{i}\alpha\downarrow},\\
&H_4^\textbf{i}=U^\prime\sum_{\sigma}n_{\textbf{i},x,\sigma}n_{\textbf{i},y,-\sigma},
\end{split}\end{equation}
where $\alpha$ denotes the $d_{xz}$ or $d_{yz}$ orbital and
$U^\prime$ satisfies the constraint $U^{\prime}=U-2J$ due to the
rotational invariance.\cite{Dagotto2001} Throughout this work, the
correlation strength is taken up to the intermediate range, i.e.,
$U/\left|t_1\right| \lesssim 2$ for both undoped and doped cases,
which is believed to be proper for the pnictides SCs.\cite{Dai2012}

In Eq.~\eqref{eq2}, $H_1^\textbf{i}$ can be
transformed\cite{Sakai2004} as
\begin{equation}\label{eq3}
   e^{-\Delta\tau H_1^\textbf{i}}=\frac{1}{2}\sum_{\gamma=\pm1}
   e^{\lambda\gamma(f_{\textbf{i}\uparrow}-f_{\textbf{i}\downarrow})}
   e^{a(N_{\textbf{i}\uparrow}+N_{\textbf{i}\downarrow}) + bN_{\textbf{i}\uparrow}N_{\textbf{i}\downarrow}}
\end{equation}
with
\begin{equation}\begin{split}\label{eq4}
   f_{\textbf{i},\sigma}&=d_{\textbf{i},x,\sigma}^\dagger d_{\textbf{i},y,\sigma}+
   d_{\textbf{i},y,\sigma}^\dagger d_{\textbf{i},x,\sigma},\\
   N_{\textbf{\textbf{i}},\sigma}&=n_{\textbf{i},x,\sigma}+n_{\textbf{i},y,\sigma}-
   2n_{\textbf{i},x,\sigma}n_{\textbf{i},y,\sigma},
\end{split}\end{equation}
where $a$, $b$ and $\lambda$ are some parameters depending on Hund's
coupling $J$ and Trotter interval $\Delta\tau$, and $\gamma=\pm1$ is
the newly introduced auxiliary field.

Due to the property that
$N_{\textbf{i},\sigma}^2=N_{\textbf{i},\sigma}$, the factor
$e^{bN_{i\uparrow}N_{i\downarrow}}$ in Eq.~\eqref{eq3} can be
further decoupled into a product of single $e^{N_{i\sigma}}$-like
terms using the discrete HS transformation \cite{Hirsch1983}. Then
all the terms containing $e^{N_{i\sigma}}$, which are independent of
the introduced field $\gamma$ in Eq.~\eqref{eq3}, can be combined
with $H_2^\textbf{i}$ in Eq.~\eqref{eq2} for the ordinary CPMC
treatment. However, after this recombination, we can see that the
remaining factor in Eq.\eqref{eq3},
$e^{\lambda\gamma(f_{\textbf{i}\uparrow}-f_{\textbf{i}\downarrow})}$,
contrary to other interactions which are made up of the number
operator $n_{\textbf{i},\alpha,\sigma}$, involves some hopping-like
terms. So some adjustment must be made for this new item
$e^{\lambda\gamma(f_{\textbf{i}\uparrow}-f_{\textbf{i}\downarrow})}$.

Recalling that in the standard QMC algorithm, the matrix form of the
interaction term, such as the Hubbard repulsion $H_1^i$, always has
the form:
\begin{equation}\label{eq5}
   e^{H_1^i}=I+A,
\end{equation}
where $A$ is sparse with one element in the diagonal and $I$ is the
identity matrix. Consequently, the determinant division $\frac{\det
L(I+A)R}{\det LR}$ and the matrix inverse $(L(I+A)R)^{-1}$ can be
calculated using a fast updating
algorithm.\cite{White1989,Hanke1992}

  We find that the matrix form of
$e^{\lambda\gamma f_{\textbf{i}\sigma}}=
e^{\lambda\gamma(d_{\textbf{i},x,\sigma}^\dagger
d_{\textbf{i},y,\sigma}+\mathrm{h.c.})}$ can be cast into a similar
form as Eq.~\eqref{eq5}:
\begin{equation}
e^{\lambda\gamma(d_{\textbf{i},x,\sigma}^\dagger
d_{\textbf{i},y,\sigma}+\mathrm{h.c.})}=I+B
\end{equation}
but with $B$ having four non-zero elements
\begin{equation}
B= \begin{pmatrix}
&\ddots &&&&&\\
&& b_{mm} &\cdots &b_{mn} &&\\
&&\vdots & \ddots & \vdots &&\\
&& b_{nm} &\cdots & b_{nn} &&\\
&&&&&\ddots &\\
\end{pmatrix} ,
\end{equation}
where
$b_{mm}=b_{nn}=\frac{e^{-\lambda\gamma}+e^{\lambda\gamma}}{2}-1$,
$b_{mn}=b_{nm}=\frac{-e^{-\lambda\gamma}+e^{\lambda\gamma}}{2}$. If
we insert the unitary matrix  $UU^{-1}(I+B)UU^{-1}$ to make
$U^{-1}(I+B)U=I+B^{\prime}$ with $B^{\prime}$ the desired diagonal
form as $A$ in Eq.~\eqref{eq5}, the determinant division $\frac{\det
L(I+B)R}{\det LR}$ and the matrix inverse $(L(I+B)R)^{-1}$ can then
be written as
\begin{equation}\begin{split}\label{eq8}
&\frac{\det L(I+B)R}{\det LR}=\frac{\det L^{\prime}(I+B^{\prime})R^{\prime}}{\det LR}, \\
&(L(I+B)R)^{-1}=(L^{\prime}(I+B^{\prime})R^{\prime})^{-1} ,
\end{split}\end{equation}
where $L^{\prime}=LU$ and $R^{\prime}=U^{-1}R$. Now the standard
CPMC algorithm can be applied with the new formulas of
Eq.~\eqref{eq8}.

In order to investigate the magnetic properties, we examine the
magnetic correlations through the static magnetic structure factor
\begin{equation}
   S(k)=1/N\sum_{ij} e^{i\textbf{k}\cdot(\textbf{r}_i-\textbf{r}_j)}
\langle(n_{\textbf{i}\uparrow}-n_{\textbf{i}\downarrow})(n_{\textbf{j}\uparrow}-n_{\textbf{j}\downarrow})\rangle
,
\end{equation}
where
$n_{\textbf{i}\sigma}=n_{\textbf{i},x,\sigma}+n_{\textbf{i},y,\sigma}$.

%%----------------table-------------------------%%
\begin{table}
\caption{The possible nearest-neighbor-bond pairing basis matrices
of the two-orbital models used in our simulations
(Ref.~\onlinecite{Wan2009}). The first column is the index number,
the second and third columns list the representations and the basis
matrices $f(\textbf{k})\tau_i$. The last column shows the spin
parities where S refer to singlet and T to triplet. Note that a
nodeless $s_{\pm}$ is also listed in the first row.} \centering
\begin{tabular}[b]{cccc}
\hline\hline %inserts double horizontal lines
No.&        IR&   $f(\textbf(k))\tau_i$&   Spin \\[0.5ex]\hline
$s_{\pm}$&  $A_{1g}$&    $\cos k_x \cos k_y\tau_0$&      S \\
        2&  $A_{1g}$& $(\cos k_x+\cos k_y)\tau_0$&      S \\
        3&  $A_{1g}$& $(\cos k_x-\cos k_y)\tau_0$&      S \\
        4&  $A_{2g}$& $(\cos k_x-\cos k_y)\tau_1$&      S \\
        6&  $B_{1g}$& $(\cos k_x-\cos k_y)\tau_0$&      S \\
        7&  $B_{1g}$& $(\cos k_x+ \cos k_y)\tau_3$&      S \\
        9&  $B_{2g}$& $(\cos k_x+ \cos k_y)\tau_1$&      S \\
       10&   $E_{g}$&         $\sin k_xi\tau_2$&      S \\
       12&  $A_{2g}$&$(\cos k_x+\cos k_y)i\tau_2$&      T \\
       13&  $B_{2g}$&$(\cos k_x-\cos k_y)i\tau_2$&      T \\
       14&   $E_{g}$&          $\sin k_x\tau_0$&      T \\
       15&   $E_{g}$&          $\sin k_x\tau_3$&      T \\
       16&   $E_{g}$&          $\sin k_x\tau_1$&      T \\ [1ex]
\hline\hline
\end{tabular}
\label{pairnumber}
\end{table}

Concerning the pairing properties, the classification of
possible pairing symmetries in Ref.~\onlinecite{Wan2009} is followed
(see Table.~\ref{pairnumber}). In the multi-orbital systems, the pairing
operators have both spatial and orbital degrees of
freedom.\cite{Moreo2009b} The singlet and triplet (with projection
1) pairing operators, $\Delta_s^{\dagger }(\textbf{k})$ and
$\Delta_t^{\dagger }(\textbf{k})$, can be respectively defined as
\begin{equation}\begin{split}
&\Delta_s^{\dagger }(\textbf{k})=
\frac{1}{\sqrt{2}}f(\textbf{k})(\tau_i)_{\alpha,\beta}
(d_{\textbf{k},\alpha,\uparrow}^{\dagger}d_{-\textbf{k},\beta,\downarrow}^{\dagger}-
d_{\textbf{k},\alpha,\downarrow}^{\dagger}d_{-\textbf{k},\beta,\uparrow}^{\dagger}) , \\
\notag %%
&\Delta_t^{\dagger
}(\textbf{k})=f(\textbf{k})(\tau_i)_{\alpha,\beta}
d_{\textbf{k},\alpha,\uparrow}^{\dagger}d_{-\textbf{k},\beta,\uparrow}^{\dagger}
,
\end{split}\end{equation}
where $d_{{\textbf{k},\alpha,\sigma}}^{\dagger}$ creates an electron
in orbital $\alpha$ with momentum $\textbf{k}$ and spin $\sigma$,
and $f(\textbf{k})$ is the form factor that transforms according to
one of the irreducible representations of the symmetry group
\cite{Moreo2009b} (for concrete forms see Table.~\ref{pairnumber}),
while $\tau_i$'s are the Pauli matrices ($i=1,2,3$) or identity
matrix ($i=0$). Using the Fourier transformation, we can get the
pairing operator in coordinate space $\Delta(\textbf{i})$, and the
corresponding pairing correlation function is defined as
\begin{equation}
P(r=\left|\textbf{i}-\textbf{j}\right|)=\langle \Delta^{\dagger}(\textbf{i})\Delta(\textbf{j})\rangle.
\end{equation}

Our CPMC code is checked by comparing to the Lanczos results on
the 2$\times$2 and 3$\times$2 clusters and also to a previous
8-site cluster Lanczos simulation\cite{Moreo2009}. Our CPMC data
are completely consistent with those results.

%%the THIRD part
\section{RESULTS AND DISCUSSIONS}\label{3rd}
\subsection{Magnetic property}

\begin{figure}[tbp]
\includegraphics[scale=0.7]{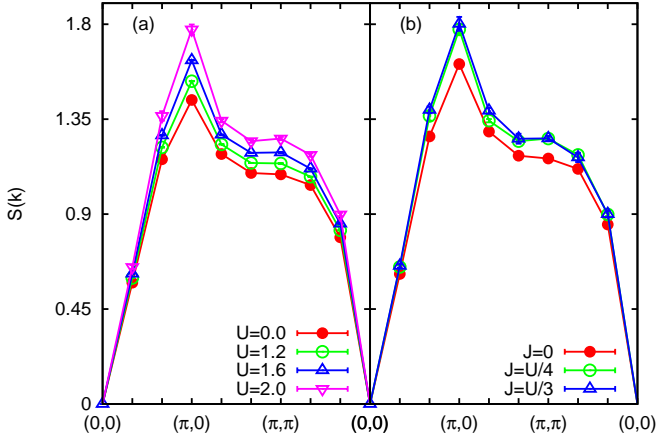}
\caption{(Color online) The static spin structure factor $S(k)$ at
half filling on 6$\times$6 cluster for (a) different on-site Coulomb
repulsions $U$ with fixed $J=0.25U$ and (b) different Hund's
couplings $J$ with $U=2.0$. } \label{skuj}
\end{figure}

\begin{figure}[tbp]
\includegraphics[scale=0.7]{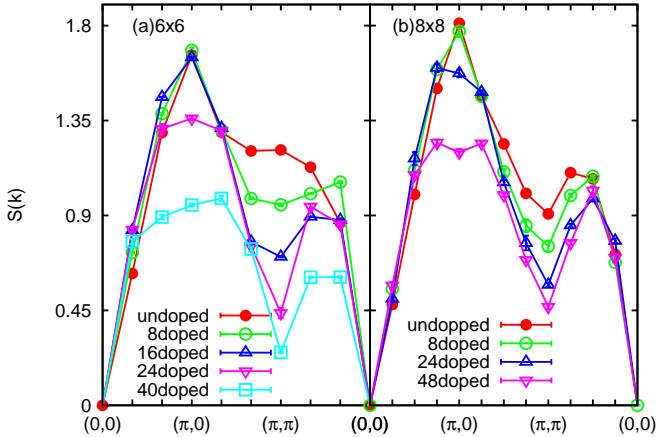}
\caption{(Color online) The effect of electron doping on the spin
structure factor with $U=1.4, J=0.25U$ on (a) 6$\times$6 and (b)
8$\times$8 lattices. The integer before doped denotes the number
of doped electrons.} \label{skde}
\end{figure}

%% undoped case: the effect of U, J on Sk
First we discuss the magnetic order in the undoped system. As shown in
Fig.~\ref{skuj}(a), the magnetic structure factor $S(k)$ is presented
at half filling (one electron per orbital) for different Coulomb
repulsions $U$ and Hund's couplings $J$ on the 6$\times$6 lattice. It
is obvious that the sharp peak at $(\pi,0)$/$(0,\pi)$ persistently exists
at various $U$ and $J$, signifying a robust $(\pi,0)$/$(0,\pi)$ magnetic order.
In addition, such a stable spin order still persists on the 8$\times$8 lattice
[see Fig.~\ref{skde}(b)]. It is worth noting that the
$(\pi,0)$/$(0,\pi)$ peak in $S(k)$ can not be viewed as a criterion
for the formation of the striped AFM order\cite{Dong2008,Paglione2010}, as we
will discuss later that another proposed magnetic order, the OM order\cite{Lorenzana2008},
also has a similar magnetic structure.

%%------------------Figures---------------------------%%
%%the figure describing the effort of U and J on sk

In Fig.~\ref{skuj}(a), we see that when increasing the Coulomb repulsion $U$,
the magnetic order is enhanced. Since the strength of the Coulomb
repulsion (in units of $\left|t_1\right|$) can be viewed as a
measurement of the electronic correlation strength, such a
$U$-induced enhancement implies the important role of electronic
correlations for the investigated magnetic order. Similarly, an
enhancement in the magnetic order is again observed when increasing
the Hund's coupling $J$ at fixed $U=2.0$ [see Fig.~\ref {skuj}(b)],
considering that $J$ favors the local magnetic moments, which also
signals possible contributions of the local moments to this magnetic
order. Within the same argument, the robust $(\pi,0)$/$(0,\pi)$ peak
at $U=0.0$ [see Fig.~\ref{skuj}(a)] indicates that the magnetic
order does not only relate to the electronic correlations and local
moments, but also to other factors, such as the FS nesting.

%----------------------------------------------------------

%----------------------------------------------------------
\begin{figure}[tbp]
\includegraphics[scale=0.7]{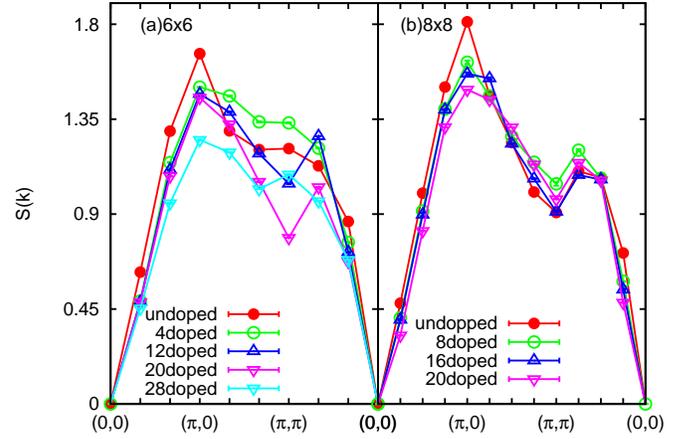}
\caption{(Color online) The effect of hole doping on the spin
structure factor with $U=1.4, J=0.25U$ on (a) 6$\times$6 and (b)
8$\times$8 lattices. The integer before doped denotes the number
of doped holes.} \label{skdh}
\end{figure}
%% the dopping effect on Sk

Next we discuss the doping effects on the magnetic order. Upon
electron doping, as shown in Fig.~\ref{skde}(a), the $(\pi,0)$ peak
seems to be unaffected compared with the undoped case intially, but
the values of $S(k)$ along the $(\pi,0)$--$(0,0)$ direction are
strongly suppressed. As a result, the magnetic order is relatively
enhanced. When more electrons are doped, the $(\pi,0)$ point starts
falling and a probable incommensurate $(\pi,0)$ magnetic structure
arises. Different from previous studies, we find that the effect of
electron doping on the magnetic order is not a monotonic suppression
and there may exist a small regime close to half filling where the
magnetic order is enhanced or at least unaffected by doping. Similar
phenomena are also observed on the 8$\times$8 lattice as shown
in Fig.~\ref{skde}(b).

\begin{figure}[tbp]
\includegraphics[scale=0.7]{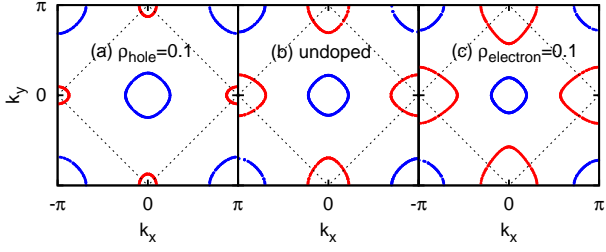}
\caption{(Color online) The Fermi surface in the extended Brillouin
zone (1 Fe per unit cell) for (a) the hole-doped, (b) undoped, and
(c) electron-doped systems with $U=0.0$, $t_1=-1.0$, $t_2=1.3$ and
$t_3=t_4=-0.85$, where the dashed lines denote the folded Brillouin
zone (2 Fe per cell) and the red (blue) curves represent the electron
(hole) Fermi pockets. } \label{fs}
\end{figure}

In the hole-doped case, however, because of the particle-hole
asymmetry of the two-orbital model, the behaviors of $S(k)$ with
doping are different. In Fig.~\ref{skdh}, the $(\pi,0)$ peak is
directly suppressed even at very low doping densities, and the
values along with the $(\pi,\pi)$--$(0,0)$ direction are ralatively
insensitive to the doping concentration. Interestingly, as reflected
in Fig.~\ref{fs}, the different behaviors of the magnetic order for
different dopants seem to be closely associated with their different
FS evolutions upon doping: With exactly the same doping density, it
is manifested that the electron pocket is notably diminished by hole
doping while that of the electron-doped system is just slightly
enlarged. On the other hand, in both cases the hole pockets almost
remain unchanged. These facts may imply that the FS nesting remains in
good condition at low electron doping while weakens at strong
electron or hole dopings. This explains why the enhancement of
magnetic order is observed only at low electron dopings. Therefore,
we propose that at least in the intermediate interaction regime the
FS nesting plays an important role in the magnetism of the
two-orbital system.

%-------------------------------------------------------
\begin{figure}[tbp]
\includegraphics[scale=0.7]{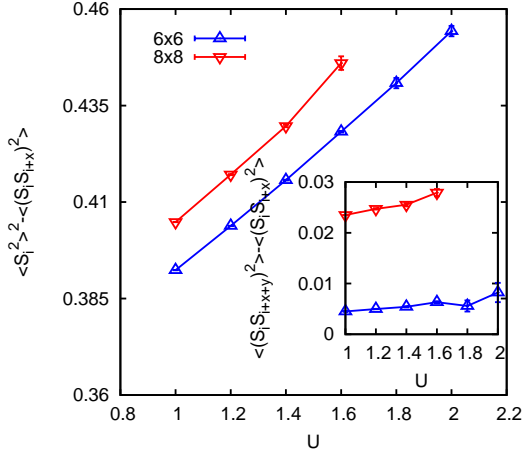}
\caption{(Color online)
$\langle\vec{S}_{\textbf{i}}^2\rangle^2-\langle(\vec{S}_{\textbf{i}}
\cdot \vec{S}_{\textbf{i}+\hat{x}})^2\rangle$ versus Coulomb
repulsion $U$ on 6$\times$6 and 8$\times$8 lattice with $J=0.25U$ at
half filling. In order to further confirm the magnetic order, the
dependence of $\langle(\vec{S}_{\textbf{i}} \cdot
\vec{S}_{\textbf{i}+\hat{x}+\hat{y}})^2\rangle-
\langle(\vec{S}_{\textbf{i}} \cdot
\vec{S}_{\textbf{i}+\hat{x}})^2\rangle$ on $U$ is illustrated in the
inset. } \label{sisj}
\end{figure}
%-------------------------------------------------------
%% the evolution of SM and OM phase when half filling
Now we analyze the competing magnetic orders of the two-orbital
model at half filling. As proposed in Ref.~\onlinecite{Lorenzana2008},
the OM order, in which the magnetic moments on nearest-neighbor sites are
at right angles, is recommended in the two-orbital model. Numerically, it
is rather difficult to distinguish the striped AFM and the OM order: both
of them have similar magnetic structure factors, negative next-nearest-neighbor
spin-spin correlations and almost-zero expectations of the nearest-neighbor
spin-spin correlations\cite{Moreo2009}. In order to identify the
competing magnetic orders at half filling, we calculate the
expectation values of the four-spin-operator
$\langle(\vec{S}_{\textbf{i}} \cdot
\vec{S}_{\textbf{i}+\hat{x}})^2\rangle$. If $U$ favors the OM order,
$\langle(\vec{S}_{\textbf{i}} \cdot
\vec{S}_{\textbf{i}+\hat{x}})^2\rangle$ should grows slower than
$\langle\vec{S}_{\textbf{i}}^2\rangle^2$. As a result,
$\langle\vec{S}_{\textbf{i}}^2\rangle^2-\langle(\vec{S}_{\textbf{i}}
\cdot \vec{S}_{\textbf{i}+\hat{x}})^2\rangle$ should increase when
increasing $U$. In Fig.~\ref{sisj}, a clear $U$-dependent
enhancement of
$\langle\vec{S}_{\textbf{i}}^2\rangle^2-\langle(\vec{S}_{\textbf{i}}
\cdot \vec{S}_{\textbf{i}+\hat{x}})^2\rangle$ is observed on both
6$\times$6 and 8$\times$8 lattices, which implies a strong tendency
for the formation of the OM order as $U$ is increased. In addition,
such a tendency becomes stronger when the lattice size is enlarged
from 6$\times$6 to 8$\times$8.

To substantiate this argument, we also calculate
$\langle(\vec{S}_{\textbf{i}} \cdot
\vec{S}_{\textbf{i}+\hat{x}+\hat{y}})^2\rangle-
\langle(\vec{S}_{\textbf{i}} \cdot
\vec{S}_{\textbf{i}+\hat{x}})^2\rangle$. Similarly, if $U$ is in
favor of the OM order, the nearest-neighbor spin-spin correlation
ought to grow slower than the next-nearest-neighbor one. Then,
$\langle(\vec{S}_{\textbf{i}} \cdot
\vec{S}_{\textbf{i}+\hat{x}+\hat{y}})^2\rangle-
\langle(\vec{S}_{\textbf{i}} \cdot
\vec{S}_{\textbf{i}+\hat{x}})^2\rangle$ should also be enhanced by
$U$, which is demonstrated by the results presented in the inset of
Fig.~\ref{sisj}.

From the above discussions, we conclude that at least in the weak to
intermediate electronic correlation regime, the magnetic order at
half filling in the two-orbital model tends to be in the OM order.
Similar conclusions are drawn from the unrestricted
Hartree-Fock\cite{Lorenzana2008} and DMRG\cite{Berg2010} studies of
the same model on other lattices.

%% next subsection
\subsection{Pairing symmetry}
%% the long range part of NN pairing channels
%--------------------------------------------------------
\begin{figure}[tbp]
\includegraphics[scale=0.65]{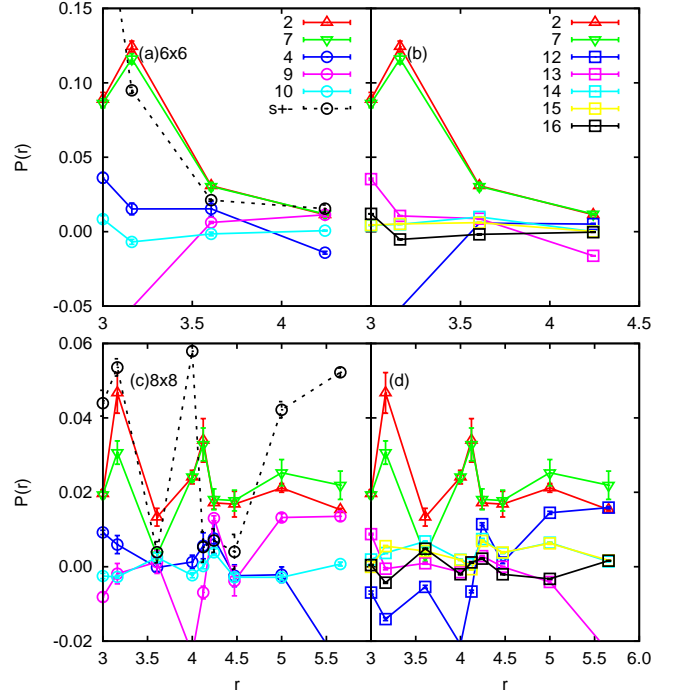}
\caption{(Color online) The non-on-site pairing correlations $P(r)$
as functions of the pairing distance $r$ in electron-doped case. (a)
and (c): The intraband singlet pairings 2, 7 v.s. all the interband
singlet pairings on 6$\times$6 and 8$\times$8 lattices,
respectively; (b) and (d): the intraband singlet pairings 2, 7 v.s.
all the triplet pairings on the same lattices. The dashed line
represents a nodeless $s_{\pm}$ pairing discussed in the context.
Here, 8 electrons are doped in the 6$\times$6 and 8$\times$8
lattices with $U=1.4$, $J=0.25U$. } \label{pde}
\end{figure}
%-------------------------------------------------------
Since the pairing symmetry is intricately related to the pairing mechanism,
it is essential to clarify the dominant pairing channel among all
the possible candidates. In this section, the long-range pairing
correlations of the possible nearest-neighbor-bond pairing
states\cite{Wan2009} and a proposed nodeless $s_{\pm}$ pairing
state\cite{Nicholson2011,Moreo2009} are discussed (see
Table.~\ref{pairnumber}), and subsequently, the effects of the
doping density $\rho$ and Coulomb repulsion $U$ on the proposed
pairing candidates are examined.

%---------------------------------------------------------
\begin{figure}[tbp]
\includegraphics[scale=0.65]{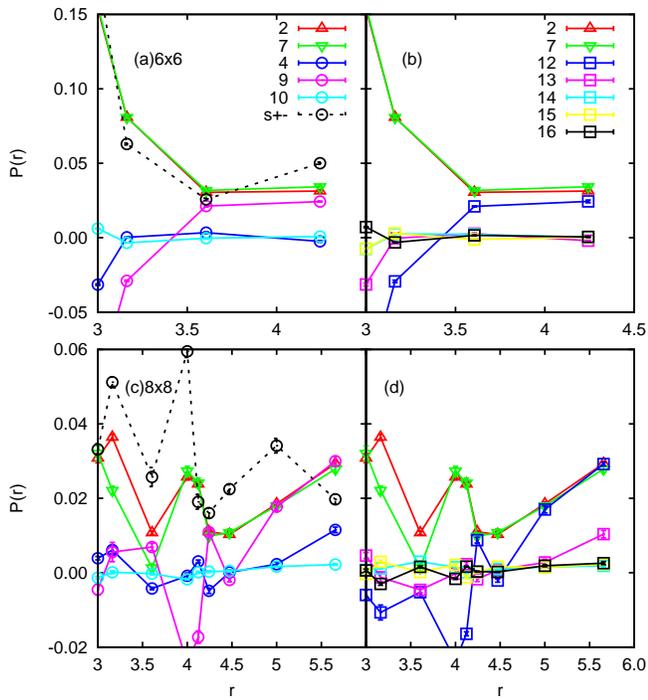}
\caption{(Color online) The non-on-site pairing correlations $P(r)$
as functions of the pairing distance $r$ in hole-doped case. (a) and
(c): The intraband singlet pairings 2, 7 v.s. all the interband
singlet pairings; (b) and (d): the intraband singlet pairing 2,7
v.s. all the triplet pairings. Similar as in Fig.~\ref{pde}, the
dash line represents the $s_{\pm}$ pairing. In this case, 4 holes
are doped in the 6$\times$6 lattice and 8 holes in the 8$\times$8
lattices with $U=1.4$, $J=0.25U$. } \label{pdh}
\end{figure}
%%-----------------------------------------------------

In Figs.~\ref{pde}(a) and \ref{pde}(b), the pairing correlations of possible
nearest-neighbor-bond pairings are shown as a function of pairing
distance $r$ when 8 electrons are doped into the 6$\times$6 system.
The pairings 2 and 7, which correspond to the spin singlet $A_{1g}$
and $B_{1g}$ intraband pairings, have the strongest amplitude at
long distances. To see more clearly, all the singlet interband and
triplet pairings are compared with pairings 2 and 7 in separated
panels (a) and (b). Since the importance of the nodeless $s_{\pm}$ pairing
with a next-nearest-neighbor-bond pairing~\cite{Nicholson2011,Moreo2009},
we also show the corresponding pairing correlation in Fig.~\ref{pde}(a).
The $s_{\pm}$ pairing also has strong long-range
pairing correlations, sometimes even stronger than that of pairings 2 and 7.
The dominance of the three competing pairings are also revealed on the
8$\times$8 lattice [see Figs.~\ref{pde}(c) and \ref{pde}(d)].

In the hole-doped case (see Fig.~\ref{pdh}), similar phenomena are
observed on both 6$\times$6 and 8$\times$8 lattices. Remarkably,
from Figs.~\ref{pde} and~\ref{pdh}, we find that the degeneracy not
only occurs between pairings 2 and 7, but also among other pairings.
For example, the singlet interband pairing 4 with $A_{2g}$ symmetry
almost has the same behaviors as the triplet pairing 13 with
$B_{2g}$ symmetry; the singlet $B_{2g}$ interband pairing 9 also
competes with the $A_{2g}$ triplet pairing 12, and so on.

\begin{figure}[tbp]
\includegraphics[scale=0.65]{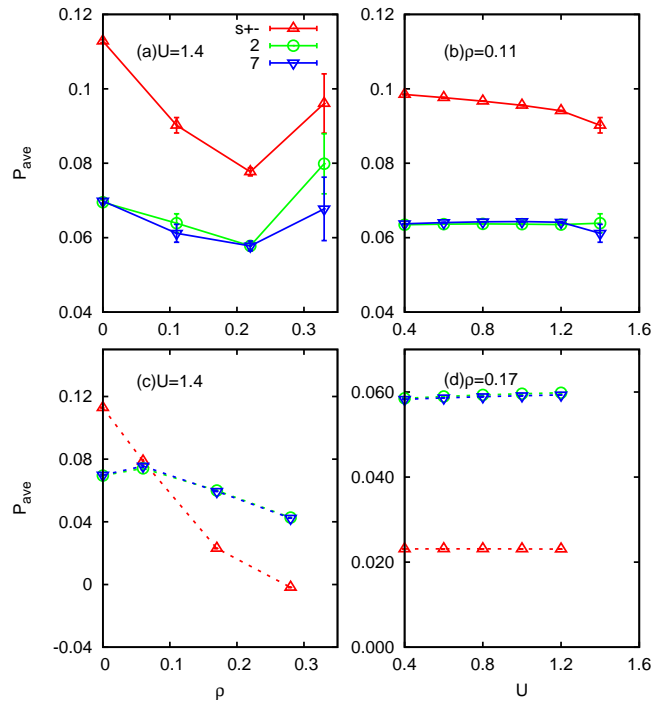}
\caption{(Color online) The average of long-range pairing
correlation $P_{\text{ave}}$ of 2, 7, and $s_{\pm}$ as functions of
the doping density $\rho$ (a) and (c), and the Coulomb repulsion $U$
(b) and (d) on the 6$\times$6 lattice. (a) and (b) correspond to the
electron-doped cases and (c) and (d) the hole-doped cases.
Here, 8 electrons are doped in (b) and 4
holes in (d). } \label{pdu}
\end{figure}

To illustrate the effect of doping density $\rho$ and Coulomb
repulsion $U$ on the proposed pairing channels, the average of
long-range pairing correlation,
$P_{\text{ave}}=\frac{1}{M}\sum_{r>3}P(r)$ with $M$ the number of
pairs, is plotted in Fig.~\ref{pdu} as functions of $\rho$ and $U$
for electron- and hole-doped cases.
%-----------------------------------------------------------
%%-------------------------------------------------------------%%
From Figs.~\ref{pdu}(a) and \ref{pdu}(c), we observe that in both the
electron- and hole-doped cases, the $s_{\pm}$ pairing, together with
pairing 2 and 7, are suppressed when increasing the doping density
$\rho$. Obviously, the $s_{\pm}$ pairing is dominant in the
electron-doped case, whereas in the hole-doped case the suppression
of the $s_{\pm}$ pairing is more drastic than that of the pairing 2
and 7, and in contrary to the electron-doped case, the pairings 2 and
7 become the dominant ones when $\rho>0.06$.

Lastly, with a fixed doping density $\rho$, we study the effect of
the Coulomb repulsion $U$ on the pairing correlations, as presented
in Figs.~\ref{pdu}(b) and \ref{pdu}(d). Overall, the effect of $U$ is much
weaker than that of doping---the pairing properties are almost
unchanged when $U$ is increased. In consistent with Figs.~\ref{pdu}
(a) and \ref{pdu}(c), the $s_{\pm}$ pairing prevails over the pairings 2 and 7
in the electron-doped case, and the latter become the leading
channels in hole doping. Thus our results demonstrate dopant-dependent
pairing symmetries of the two-orbital model.

\section{CONCLUSIONS}\label{4th}

In this paper, we have systematically studied the magnetic and
pairing properties of the two-orbital model for pnictides at half
filling and in electron- and hole-doped cases. We found that the
$(\pi,0)$/$(0,\pi)$ magnetic order is robust at half filling in the
weak to intermediate interaction regime. When increasing the Coulomb
repulsion $U$, the magnetic order is enhanced and the system tends
to be in the OM order, which is consistent with the unrestricted
Hartree-Fock and DMRG studies.\cite{Lorenzana2008,Berg2010}

When the system is doped away from half filling, the magnetic order
has different behaviors in the electron and hole dopings: It is
relatively enhanced upon the electron doping and suppressed
eventually; while in the hole-doped case, the magnetic order is
directly suppressed. Such a difference is closely relevant to
different evolutions of the FS when electrons and holes are doped in
the system---the FS nesting remains in good condition in the light
electron doping while in the hole-doped case, the electron pocket is
significantly shrunk and thus the nesting can hardly be realized.

The strong doping effects on the long-range pairing correlations
were also observed in the two-orbital model. In electron-doped case,
an $s_{\pm}$ pairing state dominates the possible
nearest-neighbor-bond pairing channels, while two nearly degenerate
intraband singlet pairing channels with $A_{1g}$ and $B_{1g}$
symmetries take over in the hole-doped case, which illustrates
a dopant-dependent pairing property of the two-orbital model.

\begin{acknowledgments}
The authors thank Adriana Moreo for useful discussions. This work was
supported by NSFC under Grants Nos. 11174072 and 91221103, and by SRFDP
under Grant No. 20104208110001.
\end{acknowledgments}

\bibliographystyle{apsrev}
\bibliography{reference}
\end{document}